\documentclass[aps,pre,preprint,showpacs]{revtex4}

\usepackage{graphicx}
\usepackage{bm}

\begin{document}

\title{Zero Temperature Properties of RNA Secondary Structures}

\author{Enzo Marinari}
\email{Enzo.Marinari@roma1.infn.it}
\author{Andrea Pagnani}
\email{Andrea.Pagnani@roma1.infn.it}
\author{Federico Ricci-Tersenghi}
\email{Federico.Ricci@roma1.infn.it}
\affiliation{Dipartimento di Fisica, SMC and UdR1 of INFM, INFN,
Universit\`a di Roma ``La Sapienza'', Piazzale Aldo Moro 2, I-00185
Roma (Italy)}

\date{\today}

\begin{abstract}
We analyze different microscopic RNA models at zero temperature.  We
discuss both the most simple model, that suffers a large degeneracy of
the ground state, and models in which the degeneracy has been remove,
in a more or less severe manner.  We calculate low-energy density of
states using a coupling perturbing method, where the ground state of a
modified Hamiltonian, that repels the original ground state, is
determined.  We evaluate scaling exponents starting from measurements
of overlaps and energy differences.  In the case of models without
accidental degeneracy of the ground state we are able to clearly
establish the existence of a glassy phase with $\theta \simeq 1/3$.
\end{abstract}


\maketitle

\section{Introduction}

RNA plays a fundamental role in the biochemistry of all living
systems~\cite{BRANDEN_TOOZE}, and it is commonly believed to be at the
origin of the pre-Darwinian epoch of life~\cite{RNA_WORLD}.  Much like
for DNA, the RNA primary structure can be described in terms of
strings of the four letter alphabet composed by Adenosine, Citosine,
Guanine, Uracile (\texttt{ACGU}).  Since RNA is usually found in the
single stranded pattern, formation of double-helix regions is
accomplished by the molecule folding back onto itself to form {\em
Watson-Crick} (WC) base pairs \texttt{G$\equiv$C} and \texttt{A$=$U},
or the slightly less stable \texttt{G$-$U} pair.  One of the most
intriguing features of RNA folded secondary structures is that in most
cases the connectivity graph is planar: This property greatly reduces
the computational efforts needed for calculating the ground state
structure.

It might be asked whether secondary structures provide an adequate
level of description for RNA real molecules~\cite{TINOCO_BUSTAMANTE}.
It is believed that secondary structure description is biologically
relevant for a number of reasons: Base pairing and base pair stacking
provide the major part of the free energy of
folding~\cite{VIENNA_PACKAGE}, secondary structures have been used
successfully by biologists in the interpretation of RNA function and
activity~\cite{RNA_WORLD}, and because structures are conserved in
evolutionary phylogeny.  At the same time computer scientists find
this level of description rather appealing since secondary structures
are discrete and therefore easy to compare.  Moreover, thanks to the
planarity condition, efficient recursive algorithms for the
computation of the native (ground state) structure are easily
implemented~\cite{NUSS_JACOB}.

Beside the genuine biological interest of RNA models, recently this
subject has raised considerable attention as an intriguing problem in
statistical mechanics of disordered systems.  The focus is now set on
the presence and the nature of a low temperature phase in ensembles of
random sequences.  In a series of recent papers
\cite{HIGGS_PRL,NOSTRO_PRL,COMM_HARTMANN,NOSTRO_REPLY,BUND_HWA_1,%
BUND_HWA_2,KMM_RNA} different authors have presented a number of
evidences (mostly numerical) supporting the existence of a transition
to a glassy phase (for a review see Ref.~\cite{HIGGS_REVIEW}).  While
a careful study of the equilibrium thermodynamics of the model
suggests a smoother than second order phase
transition~\cite{NOSTRO_PRL}, there is still much debate about the
nature of the low temperature phase, since finite size scaling
corrections are very hard to keep under control.  It has been shown
in~\cite{NOSTRO_PRL,NOSTRO_REPLY} that, at least for systems up to
1000 bases, a broad overlap distribution characterizes the low
temperature phase, but a safe extrapolation to the thermodynamic limit
from the available data is still out of
control~\cite{COMM_HARTMANN,NOSTRO_REPLY}.

Bundschuh and Hwa~\cite{BUND_HWA_1,BUND_HWA_2} have presented an
extensive study on a variety of similar RNA models supporting the
existence of a low temperature glassy phase.  They were able to
show analytically, via a two replica calculation, that weak
quenched sequence disorder is equivalent to a high temperature phase
in which all replicas are independent (molten phase), and that there
must be a finite temperature below which replicas start feeling
themselves as in a strong coupling regime.  Numerically the authors
established this glassy transition measuring the free energy cost of
imposing a pinch between two bases.  They observed that the energy of
the pinching excitation (with respect to the ground state) increases
with the sequence length following a logarithmic law (even if a power
law with small exponent was not excluded).

In this paper we study the scaling regime of the lowest energy
excitations in different models of random RNA secondary structures.
Following an idea put forward in~\cite{NOSTRO_REPLY}, we use a
perturbing method \cite{CPPS,MEZEMEZ} which has been very valuable in
the study of low temperature properties of disordered
systems~\cite{EPS_METHOD}: In the following we will call it the {\em
$\varepsilon$-coupling method}.  Very recently the same procedure has
been followed by Krzakala, M\'ezard and M\"uller~\cite{KMM_RNA}.  We
will comment on their results in the concluding section after having
presented our data.

The goal of the $\varepsilon$-coupling method is to calculate the
energy cost of typical excitations above the ground state involving a
finite fraction of the system.  As in the droplet model~\cite{DROPLET}
these energy excitations are assumed to scale as $\Delta E(L) \propto
L^\theta$, $L$ being the length of the molecule, and $\theta$ being a
relevant exponent, that we would like to determine.  It is well
possible that also the local pinching of coupled replicas imposed
in~\cite{BUND_HWA_1,BUND_HWA_2} would generate ``typical''
configurations, but we believe that our method based on a bulk
perturbation will surely do so (in this sense we find very
illuminating its application).  We apply a perturbation that is simply
a repulsion term from the ground state structure, which forces the
system to find new low-energy structures far away from the one of the
original ground state, without any other constraint.

The paper is organized as follows: in section~\ref{sec:models} we
introduce the formalism for describing the RNA secondary structure and
the different ways we have used to remove the ground state degeneracy
intrinsic to the original model~\cite{NOSTRO_PRL}.  In
section~\ref{sec:methods} we sketch the method we have applied for
calculating the low-energy spectrum of the model. We also discuss the
measurable observables.  In section~\ref{sec:results} we present our
results, focusing on the differences and similarities among the models
we have introduced.  Finally, in section~\ref{sec:conclusions}, we
summarize our findings, we compare with previous work and we comment
on further developments.

\section{Models}\label{sec:models}

The secondary structure of RNA is the set of base pairs that occur in
its three-dimensional structure. Let us define a sequence of basis as
$\mathcal{R} \equiv \{r_1,r_2,...,r_n\}$, $r_{i}$ being the $i^{th}$
base of the chain and $r_{i} \in \{A,C,G,U\}$.  A secondary structure
on $\mathcal{R}$ is now defined as a set $\mathcal{S}$ of $(i,j)$
pairs (with the convention that $1 \leq i < j \leq L$) satisfying the
following rules:
\begin{itemize}
\item $j-i \geq 4$ : this restriction permits flexibility of the chain
in its three-dimensional arrangement;
\item two different base pairs $(i,j),(i',j') \in \mathcal{S}$ if and
only if (without loss of generality we can assume that $i<i'$)
$i<j<i'<j'$, i.e.\  the pair $(i,j)$ precedes $(i',j')$, or
$i<i'<j'<j\;$, i.e.\  the pair $(i,j)$ includes $(i',j')$. This rule,
called {\em planarity condition}, excludes the occurrence of the
so-called {\em pseudo-knots}, which are very unlikely in real RNA.
\end{itemize}

We consider a simplified model for RNA folding, very similar to the
one studied in~\cite{NOSTRO_PRL}. The model is described in terms of
the Hamiltonian:
\begin{equation}
{\mathcal H} = \sum_{(i,j)\in{\mathcal S}} e_{ij} = \sum_{(i,j)}
e_{ij} \ell_{ij} \quad ,
\label{eq:ham}
\end{equation}
where $e_{ij}$ is the pairing energy between bases $i$ and $j$ and the
variable $\ell_{ij}$ takes value 1 if $(i,j)\in{\mathcal S}$ and 0
otherwise.  On a first approximation one can assume that the pairing
energies depend only on the paired bases, $e_{ij}=e(r_i,r_j)$.
Reasonable values for the energies $e(r_i,r_j)$ of the allowed base
pairs (\texttt{C-G}, \texttt{A-U} and \texttt{G-U}) at room
temperature are of $\mathcal{O}(1)$ kcal/mole~\footnote{In the models
studied here we do no try a quantitative, detailed comparison to
experimental results, and so we only select pairing energies of the
correct experimental order of magnitude.}.  One could consider other
phenomenological parameters in order to take into account the whole
complexity of a realistic energy
function~\cite{VIENNA_PACKAGE,REALISTIC_PARAM}.

We have assumed a drastic approximation in order to get a tractable
model both from a numerical and analytical point of view.  We consider
sequences made of $4$ symbols (\texttt{A}, \texttt{C}, \texttt{G} and
\texttt{U}) and we assume that only Watson-Crick base pairs may occur:
we use a strong \texttt{C-G} coupling of energy $-2$ (in arbitrary
units) and a weak \texttt{A-U} coupling of energy $-1$.  All the other
possible couplings increase the energy, so that the system avoids
these links.  One of the advantages of this model is that the role of
the disorder (encoded in the random sequence $\mathcal{R}$) is clearly
separated from that of the frustration (induced by the planarity
condition on the structure $\mathcal{S}$).

This $4$ letters model has an exponentially large ground state
degeneracy, which gives a finite $T=0$ entropy (as already found
in~\cite{NOSTRO_PRL} for the $2$ letters model).  For this reason we
refer to it as the degenerate model (the \textbf{D model}).

The large ground state degeneracy that occurs in this D model is a
pathology of a frustrated models with simple discrete interactions:
Since the couplings can take only the two negative values $-2$ and
$-1$ the same exact energetic situation can be realized in many ways.
This {\em accidental degeneracy} will probably not play a relevant
role in the physical RNA: Since real RNA energy function is far more
complex than that, ground state degeneracy is unlikely to occur.
Because of that we define two new models with modified pairing
energies, in order to remove the degeneracy.  In both models this aim
is accomplished by adding a small random perturbing term $\eta_{ij}$
to the pairing energies: $e_{ij} \to e_{ij} + \eta_{ij}$.

In the quasi-degenerate model (the \textbf{QD model}) the $\eta_{ij}$
are i.i.d.\  variables extracted from a Gaussian distribution of zero
mean $\langle \eta \rangle = 0$ and variance $\langle \eta^2 \rangle =
\eta_0^2/L$, with $\eta_0$ a small and finite constant of the order of
$0.1$: When $L\to\infty$ the pairing energies are modified of an
infinitesimal amount.  The variance is chosen such that the energies
of the ground states (which are degenerate for $\eta_0=0$) are split
over an $\mathcal{O}(\eta_0)$ range.  In this way we preserve somehow
the structure of the original energy spectrum and the sequence still
plays a key role, but the unphysical degeneracy is lifted and the
ground state is now unique.

In the non-degenerate model (the \textbf{ND model}) the variance of
the $\eta_{ij}$ variables is finite, $\langle \eta^2 \rangle =
\eta_0^2$ (we use $\eta_0 = 0.1$).  This variance induces an
$\mathcal{O}(\sqrt{L})$ splitting of the degenerated ground states,
which has to be considered as a strong reshuffling of the original
energy spectrum, since the energy gaps among levels in the original
model were of $\mathcal{O}(1)$.  The resulting energy landscape now
depends very little on the sequence.  Because of that the ND model is
very similar to the ``Gaussian disorder model'' (the \textbf{GD
model}), already discussed in~\cite{BUND_HWA_2}, where the $e_{ij}$
are i.i.d.\  Gaussian variables of zero mean and unitary variance.  In
this model the sequence plays no role.

\section{Methods}\label{sec:methods}

The $\varepsilon$-coupling method we use to calculate low-energy
excitations is the one already used in~\cite{EPS_METHOD} and
in~\cite{KMM_RNA}.  It works as follows: First of all one calculates
the ground state structure $\bm{\ell}_0 = \{\ell^{(0)}_{ij}\}$ which
minimizes $\mathcal{H}$.  Then one adds a perturbation to the
Hamiltonian, $\mathcal{H}' = \mathcal{H} - \varepsilon (1-q)$, where
$q \equiv \frac1L \sum_{ij} \ell_{ij} \ell^{(0)}_{ij} = \frac1L\,
\bm{\ell} \cdot \bm{\ell}_0$ is the overlap with the ground state
structure (note that, with this definition, the overlap is always
positive).  The perturbation term penalizes the structures which are
close to the ground state $\bm{\ell}_0$ and thus acts as a repulsive
term in the space of structures.  Finally one calculates the ground
state structure of $\mathcal{H}'$ for many values of $\varepsilon$.
Let us call these new structures $\bm{\ell}_\varepsilon$.

By definition, for any disorder realization
$\mathcal{J}=\{\mathcal{R}, \bm{\eta}\}$, both the distance
$d_\mathcal{J}(\varepsilon,L) = 1 - \frac1L\, \bm{\ell}_\varepsilon
\cdot \bm{\ell}_0$ as well as the energy difference $\Delta
E_\mathcal{J} (\varepsilon, L) = \mathcal{H}(\bm{\ell}_\varepsilon) -
\mathcal{H}(\bm{\ell}_0)$ between $\bm{\ell}_\varepsilon$ and
$\bm{\ell}_0$ are non-decreasing functions of $\varepsilon$.  Moreover
$\Delta E_\mathcal{J} (\varepsilon, L) < \varepsilon$, since the
Hamiltonian has been perturbed by a term whose absolute value is less
than $\varepsilon$, and the structures $\bm{\ell}_\varepsilon$ are
then low energy excited states of the original Hamiltonian.  We will
indicate without the $\mathcal{J}$ subscript the observables averaged
over the quenched disorder $\mathcal{J}=\{\mathcal{R}, \bm{\eta}\}$:
$d(\varepsilon,L) = \overline{d_\mathcal{J}(\varepsilon,L)}$ and
$\Delta E (\varepsilon, L) = \overline{\Delta E_\mathcal{J}
(\varepsilon, L)}$.

The algorithm for finding the new ground states of
$\varepsilon$-coupled system is exactly the same one used for the
original Hamiltonian: The repulsion from the first ground state is
included by modifying the values of the original pairing energies
$e_{ij}$.

In the thermodynamical limit structures differing by a finite $\Delta
E$ have the same intensive energy, and one could try to understand how
they are organized in the configurational space.  An interesting
question is whether, in the large $L$ limit, these structures are
extremely close together or spread over finite distances.  The answer
to this question can be given in terms of the asymptotic quantity
\begin{equation}
d_\infty(\varepsilon) = \lim_{L \to \infty} d(\varepsilon, L) \quad ,
\end{equation}
which is again a non-decreasing function of $\varepsilon$.  If
$d_\infty(\varepsilon)=0$ for any finite $\varepsilon$ then structures
with the same energy are close together, while if
$d_\infty(\varepsilon)>0$ for $\varepsilon > \varepsilon^* \sim
\mathcal{O}(1)$ then structures with the same intensive energy may
have a broad probability distribution functions of their distances and
overlaps.

In the case where $d_\infty(\varepsilon)=0$ we can derive a relation
describing the way $d(\varepsilon, L)$ vanishes.  We assume, as in the
droplet model~\cite{DROPLET}, that the energy cost of a typical
excitation involving a finite fraction of the system (i.e.\  having
finite $d$) scales with the system size as
\begin{equation}
\Delta E_{\text{typ}} \propto L^\theta \quad .
\label{eq:typ_ener}
\end{equation}
We call $\Pi(\Delta E, d, L)$ the probability distribution (over the
disorder) of excitations with energy $\Delta E$ and size $d\, L$ in
systems of size $L$.  For any fixed and finite $d \in (0,1]$, we
assume that $\Pi(\Delta E, d, L)$ has a finite weight in $\Delta E =
0$, and so, for normalization reasons, we must have $\Pi(0, d, L) =
c(d) L^{-\theta}$ for large $L$ (unless there is a delta function in
$\Delta E = 0$ as in the D model), where $c(d)$ is a smooth function
in the scaling region.

Once we add the perturbing term $-\varepsilon\, d$ to the Hamiltonian,
an excitation of size $d$ will be activated only if its energy
satisfies $\Delta E < \varepsilon\, d$.  Thus the average distance of
the new ground state structure is given by
\begin{equation}
d(\varepsilon, L) = \int_0^1 y\, \text{d}y \int_0^{\varepsilon y}
\Pi(x, y, L)\, \text{d}x = \varepsilon\, L^{-\theta} \int_0^1 y^2
c(y)\, \text{d}y \quad ,
\label{eq:ave_dist}
\end{equation}
for small $\varepsilon$ and large $L$~\footnote{The D model has $\theta=0$,
since, for a finite range of $d$ values, its $\Pi(\Delta E, d, L)$ has
a delta function in $\Delta E = 0$ that implies the presence of a gap
(energy levels are discretized), so that the integral in
Eq.~(\ref{eq:ave_dist}) is finite for any $L$ and $\varepsilon$ small
enough.}.

Then we can evaluate the $\theta$ exponent by two independent ways:
\begin{itemize}
\item from Eq.~(\ref{eq:ave_dist}), $d(\varepsilon, L) \propto
\varepsilon\, L^{-\theta}$, by measuring the average distance
$d(\varepsilon, L)$ for a fixed small $\varepsilon$ as a function of
the system size $L$;
\item from Eq.~(\ref{eq:typ_ener}), which can be equivalently rewritten
for the average energy difference as
\begin{equation}
\Delta E(d, L) \propto L^\theta \quad ,
\label{eq:ave_ener}
\end{equation}
by measuring the average energy difference for a fixed distance (not
fixed $\varepsilon$ !) as a function of the system size $L$.
\end{itemize}

\section{Results}\label{sec:results}

We study zero-temperature properties of the models described above,
i.e.\  we analyze ground states structures (GSS) of the original and of
the perturbed Hamiltonians.  We start showing the data for the $T=0$
overlap distribution in the D model (the only one with many different
ground states).  After that we present the results obtained with the
$\varepsilon$-coupling method for all the models defined in
section~\ref{sec:models}.

\subsection{The D Model}

The D model possesses an exponentially large number of GSS, which form
a set that we call $\mathcal{G}$.  In order to understand how they are
distributed in the space of structures one can calculate the
probability distribution function of the overlap, which is defined,
for any pair of structures, as $q \equiv \frac1L \sum_{ij}
\ell^{(1)}_{ij} \ell^{(2)}_{ij}$.

Unfortunately the zero-temperature entropy of the D model is too large
in order to list all the ground state structures for values of $L$
large enough to be interesting. Because of that we have added to the D
model a further constraint, suggested by observations on biological
RNA, which strongly reduces the entropy while keeping the $P(q)$ very
similar in shape to the one of the unconstrained model.  We avoid
structures where a single base pairs is surrounded by non-paired
bases, that is a structure with $\ell_{i-1,j+1}=0$, $\ell_{i,j}=1$ and
$\ell_{i+1,j-1}=0$ is forbidden.

\begin{figure}
\includegraphics[width=0.8\columnwidth]{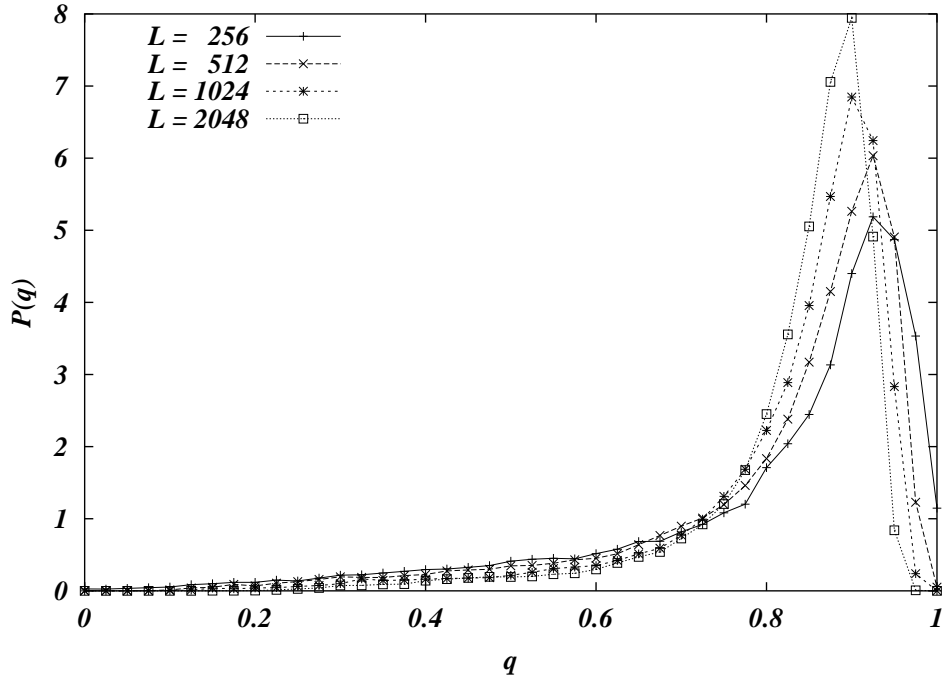}
\caption{Probability distribution of overlaps between all the pairs of
ground state structures in the degenerate D model.}
\label{pq}
\end{figure}

The overlap distribution, averaged over $1000$ samples, is shown in
figure~\ref{pq}.  Is worth noticing that the tail for small values of
$q$ is disappearing very slowly with increasing system size.  The peak
location as well as the mean overlap converge somewhere around $q
\simeq 0.87$, while the variance decreases with the system size
approximately as $\sigma^2 \propto L^{-0.45}$.  So the $P(q)$ seems to
converge, in the thermodynamic limit, to a delta function centered on
a value of $q$ smaller than $1$.  Such a value is compatible with the
observation that in a typical ground state structure the paired bases
are a finite fraction (smaller than $1$) of all bases.  Nevertheless,
as already explained in~\cite{NOSTRO_REPLY,HRT}, the triviality of the
$P(q)$ at zero temperature does not imply a trivial behavior of the
whole low-temperature phase, and so we resort to the study of
low-energy density of states.

We have calculated the ground state of the Hamiltonian $\mathcal{H}'$
for $18$ values of $\varepsilon \in [0.001,131.072]$ (equally spaced
on a logarithmic scale), and many $L$ values.  We have analyzed a
minimum of $500$ disorder realizations for the largest chain
($L=4096$), and a maximum of circa $5 \cdot 10^4$ samples for the
smallest one ($L=128$).

\begin{figure}
\includegraphics[width=0.8\columnwidth]{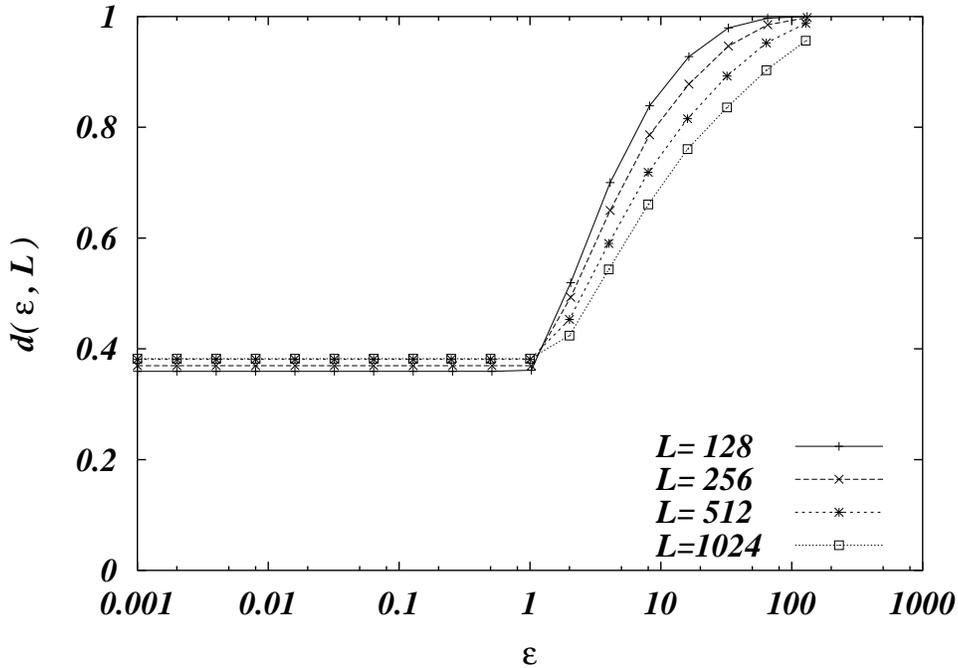}
\caption{Average distance $d(\varepsilon, L)$ versus $\varepsilon$ for
different chain lengths $L$ in the degenerate D model.}
\label{d_de}
\end{figure}

\begin{figure}
\includegraphics[width=0.8\columnwidth]{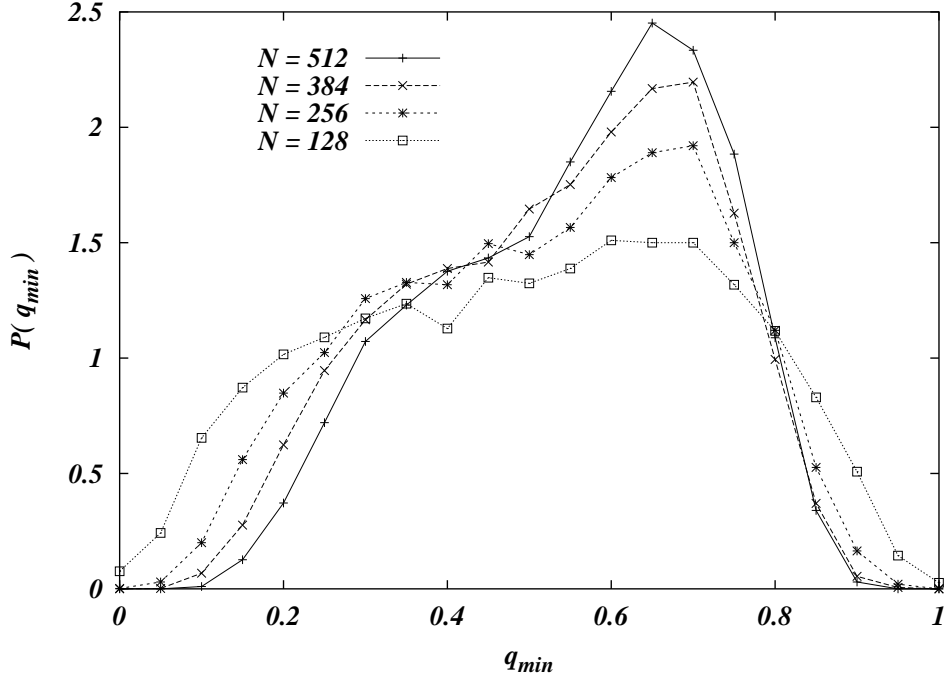}
\caption{Probability distribution $P(q_{\text{min}})$ of the minimum
overlap among any two GSS in the degenerate D model.}
\label{pqmin}
\end{figure}

The first GSS $\bm{\ell}_0$ is chosen with uniform probability in
$\mathcal{G}$ (the set of all the degenerate ground states of the D
model).  When we switch on the perturbation the new GSS will be the
one in $\mathcal{G}$ having the smallest overlap with $\bm{\ell}_0$.
Nothing else will change as long as $\varepsilon \le 1$, that is as
long as $\varepsilon$ is not large enough to make an excited state
with $\Delta E=1$ to become the new GSS.  This explains the plateau
for $\varepsilon \le 1$ in figure~\ref{d_de}, where we show the
average distance between $\bm{\ell}_0$ and $\bm{\ell}_\varepsilon$ as
a function of $\varepsilon$.  The main information we get from
figure~\ref{d_de} is the value of the plateau distance $d \simeq
0.38$, corresponding to an overlap $q = 1-d \simeq 0.62$.  This
distance can be viewed as the radius of a sphere containing the set
$\mathcal{G}$ of all the GSS. Note that the GSS are not uniformly
distributed in this sphere [otherwise the $P(q)$ would be peaked on a
much smaller overlap value], but they are very dense in the central
region and very sparse on the boundaries.  This means that if one
chooses two GSS at random they will typically be very close in the
dense region, giving a value of $q \simeq 0.87$, but if one forces the
two GSS to be as far as possible the resulting minimum overlap
$q_{\text{min}}$ will be much smaller, and will depend strongly on the
specific disorder realization [see figure~\ref{pqmin}, where we plot
its probability distribution $P(q_{\text{min}})$].

As it is made clear by the results shown in figure~\ref{d_de}, for
models with high degeneracy the perturbing method does not work
properly in the interesting region of small $\varepsilon$, giving only
information about the minimal overlap among GSS.  Now we remove the
degeneracy and analyze the other models.

\subsection{The QD Model}

\begin{figure}
\includegraphics[width=0.8\columnwidth]{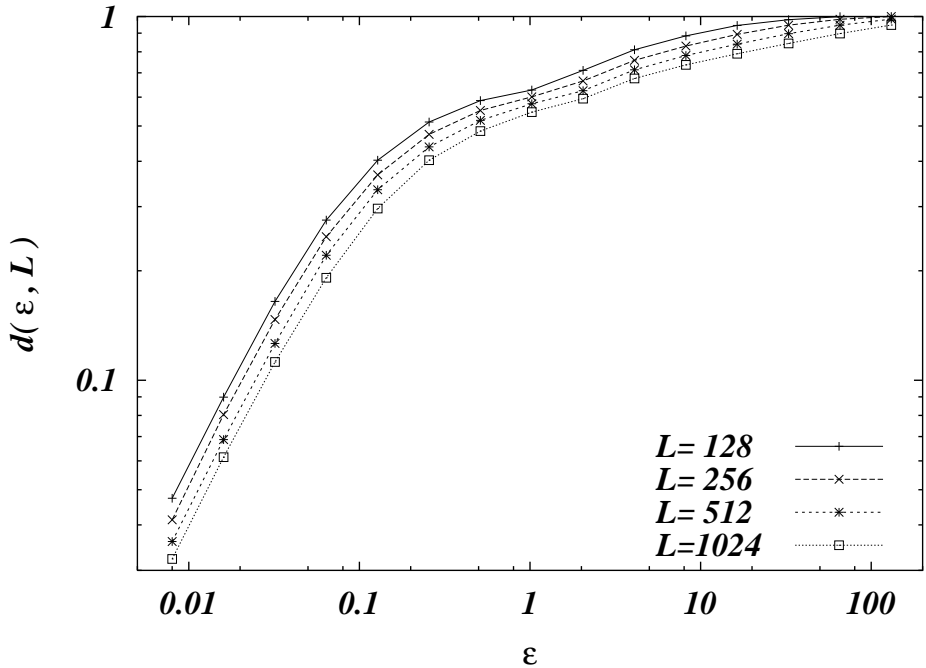}
\caption{As in Fig.~\ref{d_de} for the QD model.}
\label{qd_de}
\end{figure}

We have defined the QD model such as to keep as much as possible of
the degenerate D model, even after removal of the accidental
degeneracy.  Here we can still distinguish two different regimes (see
figure~\ref{qd_de}, where we plot the average distance $d(\varepsilon,
L)$ as a function of $\varepsilon$ for different chain lengths $L$):
For $\varepsilon > 1$ the data coincide with those for the
(unconstrained) D model, while for $\varepsilon \le 1$ they have now a
non trivial behavior.

In the interesting region of small $\varepsilon$ the scaling of the
data is very subtle and good results can be obtained either with
$\theta=0$ or with $\theta>0$.  Our finite size scaling analysis does
not allow us to reach a quantitative estimate, and we cannot
distinguish in a statistically significant way among a power law
(droplet like) scaling and a logarithmic scaling.  Further and longer
studies are needed to understand better this model.

Despite the difficulties in the data analysis, we believe that the QD
model has a large interest and relevance.  Indeed it has the great
advantage of a single non-degenerate ground state, but still the
perturbation added in order to remove the original degeneracy modifies
the energies of the structures by a quantity of order $\eta_0 \simeq
0.1$, thus keeping a large amount of information about the original
energy landscape of the degenerate D model of RNA.

\subsection{The ND Model}

In the ND model the ground state degeneracy has been removed by a
random term which strongly reshuffles the energy levels.

\begin{figure}
\includegraphics[width=0.8\columnwidth]{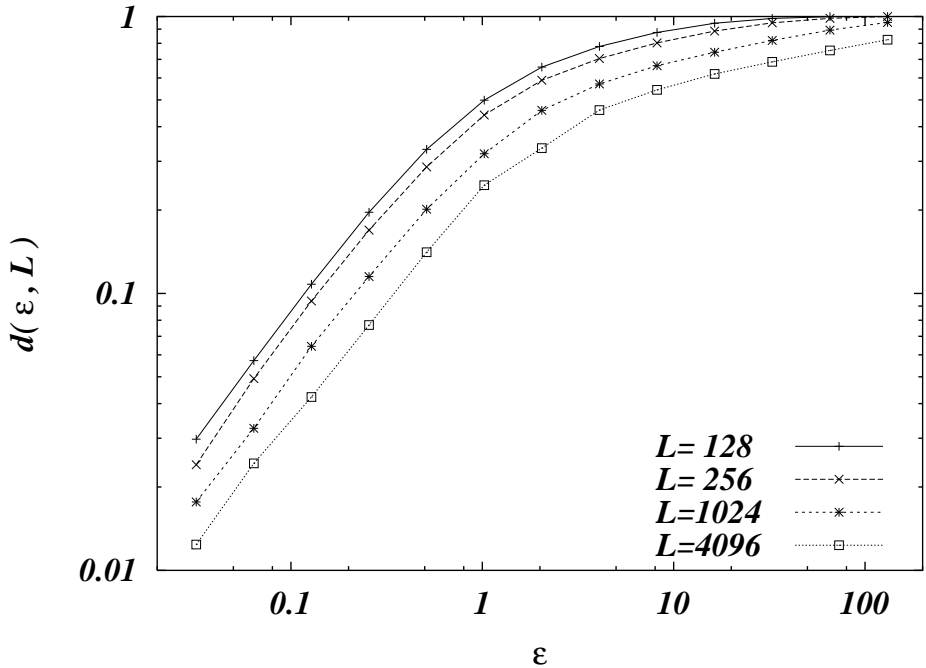}
\caption{As in Fig.~\ref{d_de} for the ND model.}
\label{nd_dens}
\end{figure}

\begin{figure}
\includegraphics[width=0.8\columnwidth]{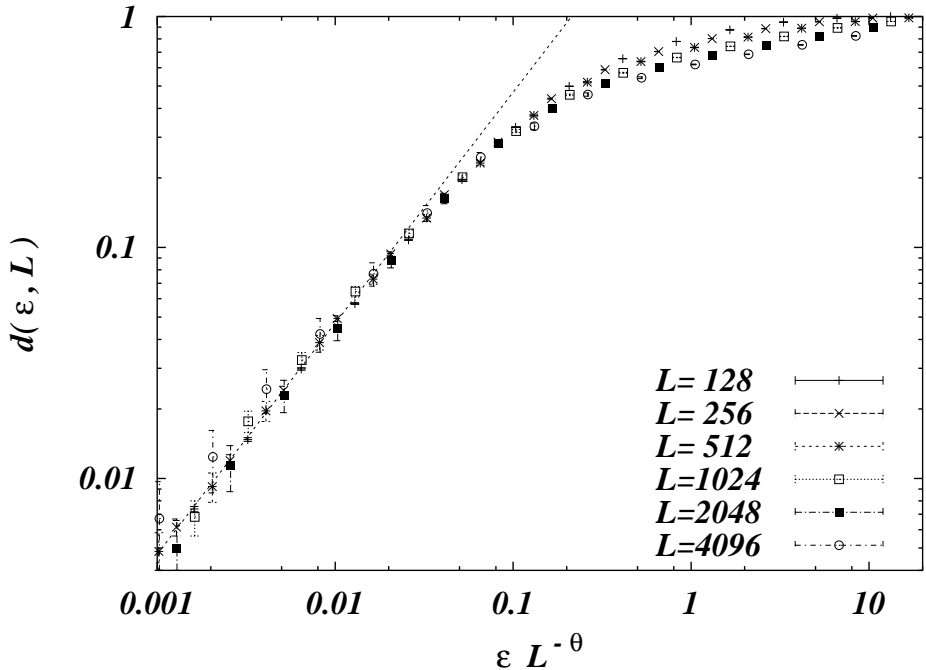}
\caption{Average distance $d(\varepsilon,L)$ as a function of the
rescaled variable $\varepsilon\, L^{-\theta}$ for different values of
$L$ in the ND model.}
\label{nd_des}
\end{figure}

In figure~\ref{nd_dens} we show the average distance
$d(\varepsilon,L)$ for some values of $\varepsilon$ and $L$ (the error
is of the order of the symbol size).  Data are now smooth functions of
$\varepsilon$, with no singular point at $\varepsilon \simeq 1$, and a
finite size scaling analysis can be performed in an easier way.

Following Eq.~(\ref{eq:ave_dist}) we have rescaled the data plotting
them versus $\varepsilon\, L^{-\theta}$.  The results are shown in
figure~\ref{nd_des}, where we have included all data points.  The best
collapse is achieved when using $\theta \simeq 0.33$.  The dotted line
has unitary slope and clearly shows that $d(\varepsilon,L) \propto
\varepsilon$ for small $\varepsilon$ and any fixed $L$.  We notice
that we are looking for a finite size scaling that works well only up
to a given value of $\varepsilon\,L^{-\theta}$: The method we are
using is based on the idea of having a ``small'' perturbation that
acts as a probe. Very large values of $\varepsilon$ drastically change
the Hamiltonian and the energy landscape. It is interesting to note
that the scaling seems to work well up to $d\simeq 0.4$, that is the
radius of the ball of degenerate ground states in the original D
model.

The physical interpretation of this result is the following.  For any
fixed $\varepsilon$ the scaling variable $\varepsilon\, L^{-\theta}$
vanishes in the thermodynamical limit, which implies that the
unperturbed ground state is stable against this kind of perturbation.
In order to have a different and more complex behavior one should
perturb the original Hamiltonian by a term whose amplitude increases
with the size at least as fast as $L^\theta$.  The energy scale
$L^\theta$ can thus be interpreted as the energy cost for reaching the
first excited state.

\begin{figure}
\includegraphics[width=0.8\columnwidth]{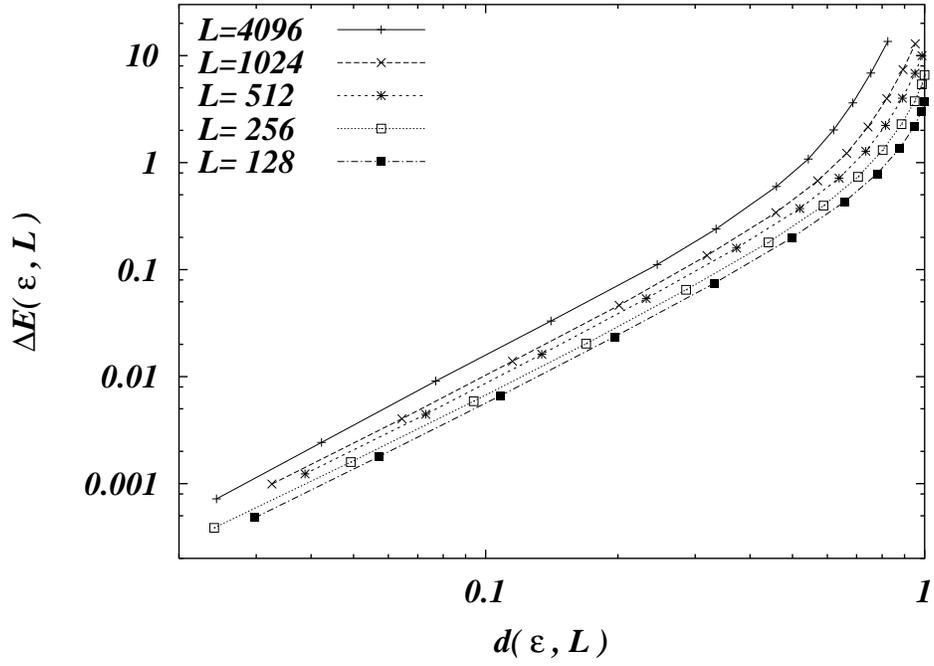}
\caption{The average energy difference $\Delta E$ between the
unperturbed and the perturbed ground state structures as a function of
their average distance $d$ for different $L$ values in the ND model.}
\label{nd_edns}
\end{figure}

\begin{figure}
\includegraphics[width=0.8\columnwidth]{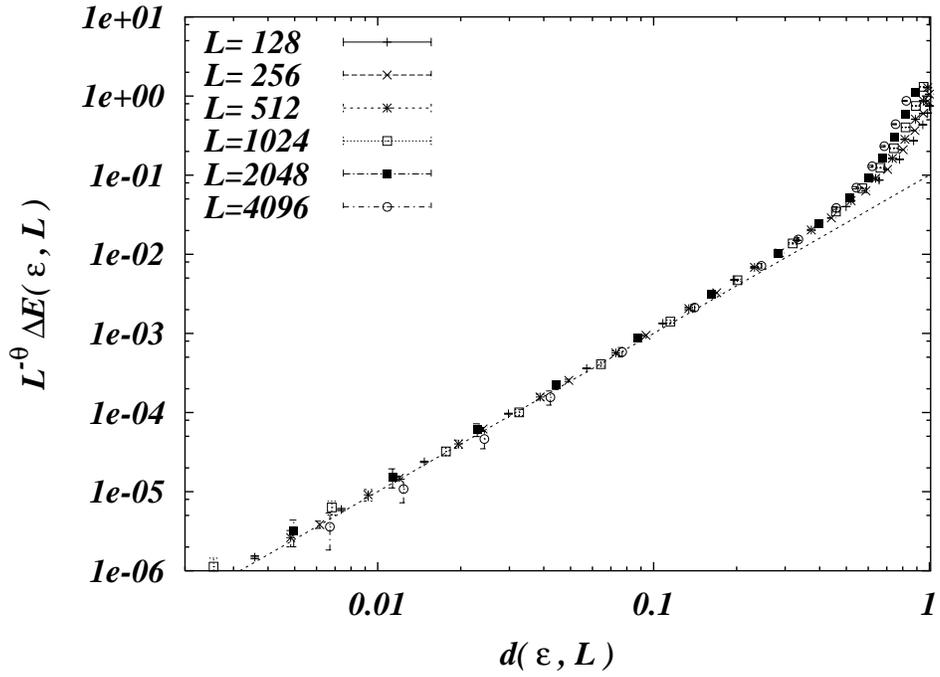}
\caption{Data of Fig.~\ref{nd_edns} rescaled according to
Eq.~(\ref{eq:ave_ener}).}
\label{nd_eds}
\end{figure}

A still clearer picture of this phenomenon is given in
figure~\ref{nd_edns}, where we plot the average energy difference
$\Delta E(\varepsilon, L)$ as a function of the average distance
$d(\varepsilon, L)$ between the unperturbed and the perturbed ground
states.  It is evident that, for any fixed distance, the energy
difference is growing with the system size, according to the argument
given above.  We have rescaled the data following
equation~(\ref{eq:ave_ener}) and the results are shown in
figure~\ref{nd_eds}.  Again the best collapse is achieved for $\theta
\simeq 0.33$, in perfect agreement with the previous analysis.  Also
here the scaling region extends over distances up to $d \simeq 0.5$.
For small distances the average energy increases as $\Delta E \propto
d^2$ (see the dotted line in figure~\ref{nd_eds}).  The same behavior
has been observed also in the QD model and we will present a simple
explanation in the next subsection, dedicated to the GD model.

Since the value of the $\theta$ exponent is small we also tried to fit
the data under the assumption $\Delta E_{\text{typ}} \propto (\log
L)^a$, that is $\theta=0$.  This behavior is suggested by mean-field
solutions of disordered models~\cite{MPV} and more particularly from
previous findings on similar models~\cite{NOSTRO_PRL,BUND_HWA_1}.  The
conclusion is that a logarithmic fit with $a = 1.85 \pm 0.15$ still
works rather well, but definitely the power law fit with $\theta =
0.33 \pm 0.01$ is more accurate and always has a smaller $\chi^2$
value.

\subsection{The GD Model}

Under the application of the $\varepsilon$-coupling perturbation, the
GD model behaves very similarly to the ND model in that:
\begin{itemize}
\item All the data perfectly collapse with $\theta \simeq 1/3$;
\item For small $\varepsilon$, $d \propto \varepsilon$;
\item For small $d$, $\Delta E \propto d^2$.
\end{itemize}
Moreover the results listed here do not depend on the presence of the
constraint $\ell_{ij}=0$ for $|i-j|<4$ (see section~\ref{sec:models}).
These findings imply that we have at hand a very simplified model
which share the same phenomenology with more realistic models and
which is more amenable to an analytical treatment.  For sake of
clarity we recall its definition: We have an Hamiltonian of the form
given in equation~(\ref{eq:ham}), where the pairing energies
$e_{ij}=e_{ji}$ are $L(L-1)/2$ independent Gaussian variables with
zero mean and unitary variance, and the $\ell_{ij}$ satisfy the
planarity condition.

Within this model it is easier, for example, to understand the
behavior $\Delta E \propto d^2$ for small $d$.  First of all we
observe from numerical simulations that in a typical GSS $\bm\ell_0$
the fraction of paired bases is $f<1$ and the distribution of the
pairing energies $e_{ij}$ of the active links, the ones with
$\ell^{(0)}_{ij}=1$, can be very well approximated by a Gaussian of
negative mean and finite width (the distribution is truncated since
positive pairing energies are forbidden in the GSS).  Let us call the
distribution of the pairing energies absolute values $P_e(e)$.  The
only property we need for the proof is a {\em finite} weight in zero,
$P_e(0)>0$, and this is the case for the GD model (and also for the ND
model).

Now we construct a sequence of structures $\bm\ell_k$ such that
$\Delta E \propto d^2$ for small $d$.  $\bm\ell_k$ is obtained from
$\bm\ell_0$ removing the $k$ weakest links, i.e.\  those with the
smallest (in absolute value) pairing energies.  So the distance
between $\bm\ell_0$ and $\bm\ell_k$ is $d=k/L$ and the energy
difference $\Delta E$ is the sum of the smallest (in absolute value)
$k$ pairing energies.  For large $L$ we can write
\begin{equation}
\Delta E = \int_0^u P_e(e')\, e'\, \text{d}e' = P_e(0) \frac{u^2}{2}
\quad ,
\end{equation}
where the last equality only holds for small $u$.  The upper
integration limit $u$ is chosen such that $k$ pairing energies, or
equivalently a fraction $2k/(fL)$ of pairing energies, are summed,
that is
\begin{equation}
\frac{2k}{fL} = \frac{2d}{f} = \int_0^u P_e(e')\, \text{d}e' = P_e(0)
u \quad ,
\end{equation}
where the last equation is valid for small $u$.  Combining the above
equations we obtain $\Delta E \propto d^2$.

Since we have chosen a sequence of structures which are not guaranteed
to have the lowest possible energies, we can only argue that $\Delta E
\propto d^\alpha$ with $\alpha \ge 2$.  Nevertheless from numerical
simulations the exponent turns out to be exactly $2$ (see
figure~\ref{nd_eds}).

In this very simplified GD model we can make one more analytical
prediction, regarding the fraction of paired bases in the GSS.  The
number of planar structures with a fraction $f$ of paired bases can be
easily calculated with the help of generating functions and turns out
to be given by $\exp[L s(f)]$, with an intensive entropy
\begin{equation}
s(f) = -f \log f - (1-f) \log(1-f) + f \log 2 \quad .
\end{equation}
$s(f)$ has a maximum for $f=\frac23$, with $s(\frac23)=\log 3$.

Let us now fix $f$ and see how the energies of the $\exp[L s(f)]$
structures are distributed.  They look random, but actually, since the
independent Gaussian random variables are only $L(L-1)/2$, there must
be many correlations among them.  Since any of these energies is the
sum of $fL/2$ random Gaussian pairing energies, we make the
approximation that the distribution of structure energies is also
Gaussian with a variance proportional to $fL$, i.e.\  $\mathcal{P}(E)
\propto \exp[-E^2/(b\,fL)]$.  The evaluation of the coefficient $b$ is
out of our present scopes.  Given, for any fixed $f$, the number of
structures and the distribution of the energies, we can estimate the
most probable lowest energy $E_{\text{min}}(f)$ through
\begin{equation}
e^{-L s(f)} = \int_{-\infty}^{E_{\text{min}}} \mathcal{P}(E') \,
\text{d}E' \simeq e^{-E_{\text{min}}^2/(bfL)} \quad ,
\end{equation}
where the last equality holds because $E_{\text{min}}$ is negative and
large. The above equation implies $E_{\text{min}}(f)=-L\sqrt{bfs(f)}$.
In order to find the fraction of paired bases corresponding to the GSS
one has to minimize $E_{\text{min}}(f)$, or equivalently maximize $f
s(f)$, over $f$.  Such an extremum is achieved for $f=0.86$ to be
compared with the fraction of paired bases found numerically
$f=0.856$.  The rather small discrepancy tells us that the
approximation made on the form of the structure energies distribution
$\mathcal{P}(E)$ is not so bad.

\begin{figure}
\includegraphics[width=0.8\columnwidth]{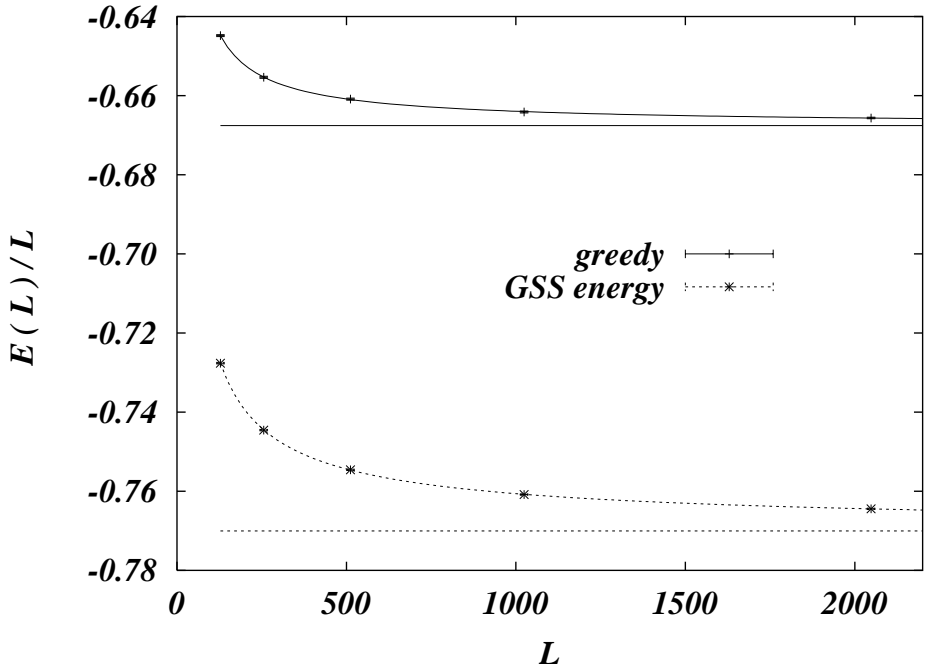}
\caption{As a function of the system size $L$ we compare the ground
state energy of the GD model (below) with the one reached by a simple
greedy algorithm (above).  The horizontal lines correspond to the
infinite size extrapolations.}
\label{greedy}
\end{figure}

The apparent correctness of such a simple approximation could suggest
that the GD model has a trivial energy landscape.  We have checked for
this possibility with the following method: In a trivial energy
landscape any reasonably smart greedy algorithm should be able to
reach, or at least to closely approach, the ground state energy.  We
have used a greedy algorithm which builds up the structure in the
following way: It starts with a structure with no links, at each step
it chooses the lowest negative pairing energy (largest in absolute
value) among the set of those allowed by the planarity condition, and
adds the corresponding link to the growing structure.  Using this
greedy algorithm we can reach the energies shown in
figure~\ref{greedy} which are more than 10\% higher than the
corresponding ground state energies.  So, it seems that finding a
structure with low energy in linear time is not an easy task. This
suggests a complex energy landscape.  A deeper analysis is obviously
needed in order to say how much complex the energy landscape of the GD
model is.

\section{Conclusions}\label{sec:conclusions}

Our results allow to describe a clear and simple physical picture for
the RNA-inspired models studied here.  All of them possess a glassy
phase at low enough temperatures (since we have analyzed very low
energy density of states we cannot make predictions on the location of
the critical temperature).  We can claim that our study clearly
selects a positive $\theta$ exponent value for all cases but for the
quasi degenerate QD model, where our numerical results are not precise
enough to allow us quantitative statements.

At variance with the results of~\cite{BUND_HWA_1,BUND_HWA_2} we find
that in the non-degenerate ND model the broken phase does not look
marginal, but a standard droplet glassy phase with $\theta \simeq 1/3
> 0$.  Our way of analyzing the data allows us to exclude (with good
confidence) a simple logarithmic divergence of the energy difference
between ground state and excited states.  On this issue we agree with
the results of~\cite{KMM_RNA}: The difference in the estimate we give
for $\theta$, as compared to the $\theta \simeq 0.23$ of
\cite{KMM_RNA}, is probably due to the different fitting procedure,
and to the number of free parameters used in the fit.

The $\theta$ exponent we find is perfectly compatible with that for
directed polymers in random media in 1+1 dimensions~\cite{MEZEMEZ},
$\theta_{DPRM}=\frac13$.  Since the two models have some similarities
this relation could indeed hide a deep connection.

The degenerate D model and the quasi degenerate QD model we have
defined above are maybe the less trivial and the most intriguing from
the theoretical point of view.  Unfortunately we were not able to
determine accurately the asymptotic scaling behavior in the latter.

It is probable, on the contrary, that the most part of the analytic
developments will be obtained for the Gaussian disorder GD model, that
is by far the simplest among all the models with a non-trivial
behavior.

\begin{acknowledgments}
We thank T. Hwa, F. Krzakala, M. M\'ezard, M. M\"uller and G. Parisi
for a number of interesting conversations.
\end{acknowledgments}


\begin{thebibliography}{99}

\bibitem{BRANDEN_TOOZE} C. Branden and J. Tooze, \textit{Introduction
to Protein Structure} (Garland Publishing, New York, 1991).

\bibitem{RNA_WORLD} R.F. Gesteland and J.F. Atkins, \textit{The RNA
World: the Nature of Modern RNA Suggests a Prebiotic RNA World} (Cold
Spring Harbor Laboratory Press, New York, 1993).

\bibitem{TINOCO_BUSTAMANTE} I. Tinoco Jr. and C. Bustamante,
J. Mol. Biol. \textbf{293}, 271 (1999).

\bibitem{VIENNA_PACKAGE} I.L. Hofacker, W. Fontana, P.F. Stadler,
S. Bonhoeffer, M. Tacker, and P. Schuster, Monatshefte f. Chemie
\textbf{125}, 167 (1994).

\bibitem{NUSS_JACOB} R. Nussinov and A.B. Jacobson,
Proc. Nat. Acad. Sci. USA \textbf{77}, 6309 (1980).

\bibitem{HIGGS_PRL} P.G. Higgs, Phys. Rev. Lett. \textbf{76}, 704
(1996).

\bibitem{NOSTRO_PRL} A. Pagnani, G. Parisi, and F. Ricci-Tersenghi,
Phys. Rev. Lett. \textbf{84}, 2026 (2000).

\bibitem{COMM_HARTMANN} A.K. Hartmann, Phys. Rev. Lett. \textbf{86},
1382 (2001).

\bibitem{NOSTRO_REPLY} A. Pagnani, G. Parisi, and F. Ricci-Tersenghi,
Phys. Rev. Lett. \textbf{86}, 1383 (2001).

\bibitem{BUND_HWA_1} R. Bundschuh and T. Hwa, e-print
\texttt{cond-mat/0106029}.

\bibitem{BUND_HWA_2} R. Bundschuh and T. Hwa, e-print
\texttt{cond-mat/0107210}.

\bibitem{KMM_RNA} F. Krzakala, M. M\'ezard, and M. M\"uller, e-print
\texttt{cond-mat/0108374}.

\bibitem{HIGGS_REVIEW} P.G. Higgs, Quart. Rev. Biophys. \textbf{33},
199 (2000).

\bibitem{CPPS} S. Caracciolo, G. Parisi, S. Patarnello, and
N. Sourlas, Europhys. Lett. \textbf{11}, 783 (1990); 

\bibitem{MEZEMEZ} M. M\'ezard,
J. Physique \textbf{51}, 1831 (1990).

\bibitem{EPS_METHOD} F. Krzakala and O.C. Martin,
Phys. Rev. Lett. \textbf{85}, 3013 (2000); M. Palassini and
A.P. Young, Phys. Rev. Lett. \textbf{85}, 3017 (2000); E. Marinari and
G. Parisi, Phys. Rev. B \textbf{62}, 11677 (2000);
Phys. Rev. Lett. \textbf{86}, 3887 (2001).

\bibitem{DROPLET} W.L. McMillan, J. Phys. C \textbf{17}, 3179 (1984);
A.J. Bray and M.A. Moore, Phys. Rev. Lett. \textbf{58}, 57 (1986);
D.S. Fisher and D.A. Huse, Phys. Rev. B \textbf{38}, 373 (1988).

\bibitem{REALISTIC_PARAM} D.H. Mathews, J. Sabina, M. Zucker, and
H. Turner, J. Mol. Biol. \textbf{288}, 911 (1999).

\bibitem{HRT} A.K. Hartmann and F. Ricci-Tersenghi, e-print
\texttt{cond-mat/0108307}.

\bibitem{MPV} M. M\'ezard, G. Parisi, and M.A. Virasoro, \textit{Spin
Glass Theory and Beyond} (World Scientific, Singapore, 1986).

\end{thebibliography}
\end{document}